\documentstyle[12pt]{article}
\begin{document}
\setlength{\baselineskip}{0.27in}

\newcommand{\kp}{k_{\perp}}
\newcommand{\La}{\Lambda_{QCD}}
\newcommand{\al}{\alpha_s}
\newcommand{\beq}{\begin{equation}}
\newcommand{\eeq}{\end{equation}}
\newcommand{\beqa}{\begin{eqnarray}}
\newcommand{\eeqa}{\end{eqnarray}}
\newcommand{\lsim}{\begin{array}{c}\,\sim\vspace{-21pt}\\<\end{array}}
\newcommand{\gsim}{\begin{array}{c}\sim\vspace{-21pt}\\>
\end{array}}

\begin{titlepage}
{\hbox to\hsize{June 1995\hfill } }
{\hbox to\hsize{UM-TH-95-19\hfill hep-ph/xxxxxxx}}
\begin{center}
\vglue .06in
{\Large \bf Leading Power Corrections in QCD: From Renormalons to Phenomenology
 }
\\[.5in]
\begin{tabular}{c}
{\bf R. Akhoury ~~and~~ V.I. Zakharov}\\[.05in]
{\it The Randall Laboratory of Physics}\\
{\it University of Michigan}\\
{\it Ann Arbor MI 48109 }\\[.15in]
\end{tabular}
\vskip 0.25cm
{\bf Abstract}\\[-0.05in]

\begin{quote}
We consider $1/Q$ corrections to hard processes in QCD where Q is a large mass
scale, concentrating on shape variables in $e^{+}e^{-}$ annihilation. While the
evidence for such corrections can be and has been established by means of the
renormalon technique, theory can be confronted with experiment only after
clarifying the properties of the corresponding non-perturbative contribution.
We list predictions based on the universality of the $1/Q$ terms,
and compare them with the existing data. We also identify the scale of the
non-perturbative
contributions in terms of jet masses.
\end{quote}
\end{center}
\end{titlepage}
\newpage

\section{Introduction}

Perturbative QCD constitutes a well defined framework for understanding
hard processes, i.e. processes characterized by a large mass scale Q
(see for example \cite{troyan}).
While the zeroth order approximation is provided usually by the parton model
and is well defined,
the radiative corrections can bring logarithmic factors which depend on the
infrared cut off
and destabilize the theoretical predictions. To avoid this problem, one
concentrates usually
on a set of infrared safe quantities which are protected against such
contributions.
There are  powerful general theorems on infrared stability of theoretical
predictions based on
factorization of short and large distances, for a review see \cite{collins}.
In this note we will consider shape variables in $e^+e^-$ annihilation which
are infrared safe.
As an example one may keep in mind the thrust $T$ defined as
\beq
T~=~max_{\bf n}{\Sigma{({\bf p}_i{\bf n})}\over \Sigma |{\bf p}_i|}
\eeq
where ${\bf p}_i$ are the momenta of the particles produced while ${\bf n}$ is
a unit vector.

While perturbative QCD allows us to calculate thrust as a series in a small
expansion parameter
$\alpha_s(Q^2)$ \cite{thrust}:
\beq
\langle 1- T \rangle~=~0.335\alpha_s(Q^2)+1.02\alpha_s^2(Q^2)+... ~~~~,
\eeq
analysis of the experimental data at least at moderate $Q^2$ indicates also the
presence of
$1/Q$ corrections \cite{barreiro,webhad}. Theoretically, such corrections are
exponentially small,
$$1/Q~\sim~exp(-b_0/2\alpha_s(Q^2))~,$$
where $b_0$ is the first coefficient in the $\beta$-function, and are clearly
beyond the reach
of a purely perturbative approach. The only consisitent way to deal with the
power corrections
\cite{svz} is provided by the operator product expansion (OPE). However, this
technique applies
only to a very selective set of physical quantities like the total cross
section.

Nevertheless, most a recently QCD-based phenomenology of the $1/Q$ corrections
has been emerging
\cite{cs,webber,ks1,dw,az,ks2}. It is worth emphasizing that the very existence
of the $1/Q$
corrections can be readily understood. Indeed, consider the emission of a soft
gluon with energy
$\omega \sim \Lambda_{QCD}$. The corresponding contribution to the thrust is of
order:
\beq
1-T_{soft}~\sim~\int_0^{\La}
{{\omega\over Q} {d\omega\over \omega} \al (\La ^2)~\sim~
{\La \over Q} }
\eeq
where the first factor in the integrand comes from the definition of the
thrust,
$d\omega/\omega$ is the standard factor for emission of a soft gluon, and the
running coupling
$\al (\La ^2)$ is of order unit for a soft gluon. Although $T_{soft}$ cannot be
calculated reliably
in perturbation theory, the presence of the $1/Q$ corrections is obvious. In
reality, the existence
of the $1/Q$ corrections has been established \cite{cs,bigi,mw,webber} not
via such simple
estimates but rather by means of infrared renormalons \cite{thooft}.
Renormalons, which are a particular set of perturbative graphs, allow us to
clarify
the convergence properties of the expansions like (2) and estimate
their uncertainty as powers of $1/Q$.

While the conclusion on the very existence of $1/Q$ corrections seems to be
guaranteed,
to develop a
phenomenology of such corrections one needs means to relate them in various
processes and to fix the
overall scale. These are the issues central to the present note, which is
considered to be a
continuation of the analysis started in \cite{az}.

It might be worth emphasizing that renormalons {\it per se} cannot be a basis
for such a
phenomenology. Indeed, to apply renormalons literally one has to calculate all
the terms
in the expansion (2) until they start to rise in the fashion prescribed by the
renormalons.
The power-like terms appear then as an uncertainty of the asymptotic expansion
and are dependent on the procedure for subtracting the perturbative
contributions. The
ambiguities of such a procedure are spelled out in Ref \cite{mueller1} and it
has never been
tried so far.

The attempts to develop the phenomenology of $1/Q$ terms are based therefore on
a mixture of
theoretical inputs and heuristic arguments.
The first estimation of $1/Q$ terms in shape variables
\cite{webber} relied on the version of the renormalon technique which replaces
the renormalons by
terms non analytical in the gluon mass  squared, $\mu^2$ \cite{bigi,bbz}.
This procedure can
be thought of only in the lowest order in $\al$. This is also true for the
modification of the
technique which assumes freezing of the running coupling at some scale
\cite{dw}. The analysis of the
data indicated that experimentally $1/Q$ terms are proportional to the $\mu/Q$
corrections found
theoretically \cite{webber}. Another line of development \cite{cs,ks1}
is to evaluate the
renormalon contributions
to a given process to all orders and to develop an operator analysis based on
an effective theory
arising in the eikonal approximation \cite{collins}.

In our previous paper \cite{az} we argued that terms of all orders in the large
coupling
$\al (\La^2)$ factorize in the $1/Q$ corrections to various shape variables
into a universal factor. We argued furthermore that non-perturbative
contributions share this
property of factorization since they are also associated with distances much
larger then $1/Q$. As a
result, one can both substantiate the relations among the $1/Q$ terms in the
shape variables
\cite{webber} and try to develop a machinery for deriving furthur experimental
consequences from QCD.
Similar relations for soft perturbative parts are
derived in Ref \cite{ks2}; no experimental consequences are claimed however
because of the
reservations for unknown non-perturbative effects.

In this paper we first consider the renormalon technique versus the general
operator product
expansion in case of the total cross section when both approaches apply. We
check on this
example the hypotheses concerning the non-perturbative terms made in \cite{az}.
We give a list of
experimental consequences which follow from the universality of $1/Q$
corrections, non-perturbative
contributions including. Partly the predictions (or their variations)
were discussed earlier in \cite{webber,dw,az}. Finally, we
identify the overall scale of the $1/Q$ corrections in terms of the
non-perturbative contribution to
jet masses which arise in the Feynman-Field type models. The considerations
of this paper are restricted to the kinematic configuration where the
dominant nonperturbative corrections come from the vicinity of the
2-jet limit in $e^{+}e{-}$ annihilation.

The outline of the paper is as follows. In sect. 2 we review the renormalon
technique in case
of the total cross section. In sect. 3 we turn to an analysis of the $1/Q$
corrections to certain
infrared safe observables in $e^+e^-$
annihilation into jets and list predictions which follow from the universality
of the $1/Q$
corrections. In sect. 5 we compare the predictions with existing experimental
data. In sect. 6 we
establish the  correspondence of the renormalon based picture for the power
corrections with the
phenomenological decription of the non-perturbative effects in jet physics a la
Feynman-Field.

\section {Renormalons vs. the operator product expansion}

In this section we will try to elucidate which properties of renormalons are
shared by the
non-perturbative conributions. As a test case we choose the total cross section
of
$e^+e^-$ annihilation since in this case a more general framework based on the
OPE is also
available.

The OPE applies in euclidean space-time. The polarization operator in euclidean
space,
$P(Q^2)$, is related
to an integral over the total cross section $\sigma(s)$:
\beq
P(Q^2)~=~{Q^2\over \pi}\int
{{\sigma(s)\over (s+Q^2)s}}.
\eeq
At large energies the total cross section is calculable perturbatively:
\beq
\sigma(s)~=~(parton ~model)\cdot
\left( 1+{\al(s)\over \pi}+1.409({\al(s)\over \pi})^2-
12.805({\al(s)\over \pi})^3+...\right)~~~,\label{pert}
\eeq
where the first three terms in the expansion have been calculated explicitly
see, for example, the second of reference \cite{troyan}.
Phenomenologically, $P(Q^2)$ can also be analyzed via QCD sum rules \cite{svz}
which
approximate $P(Q^2)$ at moderately large $Q^2$ as
\beq
Q^2{dP(Q^2)\over dQ^2}~\approx~(parton ~model)\cdot
\left(1+{\al(Q^2)\over \pi}+{\langle \alpha_s(G_{\mu\nu}^a)^2\rangle \over
Q^4}{\pi\over 6}
+...\right),\label{ope}
\eeq
where $G^a_{\mu\nu}$ is the gluon field strength tensor and the vacuum
expectation value
of $(G^a_{\mu\nu})^2$ incorporates non-perturbative effects. The theoretical
framework behind
(6) is the OPE for the T-product of two electromagnetic currents while the main
hypothesis is
that the power like corrections are responsible for violation of the asymptotic
freedom
once one starts to descend from very high $Q^2$ and to approach $Q^2\sim
1~GeV^2$. The cross section which corresponds literally to (\ref{ope})
coincides with (\ref{pert}) as far as $\alpha_s$ correction is concerned
and a term proportional to $\delta '(Q^2)$.

Now, the renormalons fall so to say in between the two representations (5) and
(6)
where the former apparently does not contain power like corrections while
the latter emphasizes the $Q^{-4}$ terms. Indeed, on one hand, renormalons are
particular
perturbative graphs with $n$ insertions of the vacuum polarization into a gluon
line.
Therefore, they are included into the expansion (5). For large
$n$ and a one-term $\beta$-function the renormalon contribution is proportional
to
\beq
P(Q^2)_{renorm}~\sim~\al^n(Q^2){1\over Q^4}
\int_0^{\sim Q^2} d^4k
\left(ln{Q^2\over k^2}\right)^nb_0^n~\sim~
\al^n(Q^2)n!({b_0\over 2})^n
\eeq
where we have retained only the contribution of the extremum of the integrand
at
\beq
k^2_{eff}~\sim~Q^2exp(-n/2)
\eeq
which is reponsible for the $n!$ growth at large $n$.

On the other hand, one can argue that the perturbative calculation is of no
relevance
once $n>N_{cr}$ where the $N_{cr}$ is defined by the condition
$(k^2_{eff})_{N_{cr}}
\sim \La^2$. Indeed, once we approach the Landau pole nonperturbative effects
are expected to
become important. The renormalon contribution at $N_{cr}$ is of order
\beq
P(Q^2)_{N_{cr}}~\sim~{\La^4\over Q^4}
\eeq
and exhibits in this way the power like corrections, shown in eq. (6). These
two facets of
renormalons reveal  mixing of perturbative and non-perturbative contributions
so that only their sum
is uniquely determined, as is emphasized in Refs \cite{david,mueller2} (for
discussion see also
\cite{svz,novikov,vz}).

Imagine now that we would like to build up a phenomenology of the power like
corrections based
on renormalons. A straightforward logical way would be to evaluate all the
perturbative terms
until they start to blow up because of the $n!$ behaviour (7); define in some
way the perturbative
series and postulate the existence of non-perturbative terms which would
compensate for the
arbitrariness in defining the perturbative part. From the practical point of
view, such a program
would be difficult to implement since, for example, one of he most advanced
calculations
of the perturbative expansion (5) does not reproduce yet the $n!$ behaviour. In
principle, such
a procedure would reproduce the $Q^{-4}$ behaviour of the non-perturbative
contribution. The
$Q^{-4}$ corrections would be subordinated however to many terms in the
perturbative expansion and
hence insignificant. The relations among the $Q^{-4}$ corrections in various
channels would be
procedure dependent.

Instead, one can pick up the renormalons from the very beginning by rewriting
one-gluon exchange
with running coupling constant $\al(k^2)$ where $k$ is the momentum flowing
through the gluon line,
as
\beq
{1\over Q^4}\int_0^{\sim~Q}d^4k\al(k^2)~=~
{1\over Q^4}\int d\sigma\int d^4k \left({k\over \La}\right)^{-2\sigma b_0}~\sim
\left({\La \over Q}\right)^4 \int
{d\sigma \over 4-2\sigma b_0}
\eeq
and defining the integral over the pole at $\sigma=2/b_0$ as, say, its
principal value.
Such a procedure corresponds to using the expansion (6) along with
representing the matrix element $<\al (G^a_{\mu\nu})^2>$ by its perturbative
part
\cite{vz}:
\beq
\langle \al(G^a_{\mu\nu})^2\rangle_{pert}~=
{}~{3\over 2\pi^2}\La^4 \int{d\sigma\over 1-\sigma b_0/2}.
\eeq
Note that the OPE can be used for the evaluation of the renormalon graphs since
the momentum flowing
through the line is much smaller than $Q^2$, see eq. (8). Calculating,
furthermore, the renormalon
contribution to various crosssections one would reproduce the relations among
the $Q^{-4}$
corrections following from the OPE since the substitution of a particular value
of the
matrix element (11) does not affect the general relations.

Our next hypothesis would be that non-perturbative contributions are
proportional to
those of renormalons. The reason is that the both are associated with large
distances
of order $\La^{-1}$ and therefore factorize simultaneously. In the case of the
total cross section
this assumption is obviously true since both the soft-perturbative and
non-perturbative
effects are absorbed into the one and the same matrix element
$<(G^a_{\mu\nu})^2>$.
Finally, the whole machinery of the sum rules would be reproduced if one
assumes that the
non-perturbative effects enhance the renormalon contributions so that
 \beq \langle
(G^a_{\mu\nu})^2\rangle _{pert} \ll~\langle (G^a_{\mu\nu})^2 \rangle
_{non-pert}.
\eeq
The inequality (12) is understood here in the practical sense that one may
neglect at moderately
large $Q^2$ high orders in $\al(Q^2)$ as compared to the $Q^{-4}$ terms.
This assumption does not have a sound theoretical foundation as yet and should
be considered
as a heuristic one, or as motivated by the analysis of the data. It is crucial,
however, to build up
a phenomenology starting from renormalons \cite{vz}.

Let us also mention two other points concerning renormalons. First, we took it
for granted
that $\al$ in eq. (10) depends on the momentum flowing through the gluon line.
This is in fact not
obvious and should have been substantiated by a careful analysis. This is a
difficult point of the
renormalon technique. Second, renormalons are pure perturbative constructs and
they cannot
distinguish which hadronic matrix element of $(G^a_{\mu\nu})^2$ is considered.
The enhancement due
to the non-perturbative effects can depend on the hadronic state, on the other
hand. Consideration
of total cross sections is related to the vacuum epectation value of
$(G^a_{\mu\nu})^2$.

To summarize, the phenomenological success of the QCD sum
rules suggests the validity of two basic observations
concerning non-perturbative terms, namely, that (a)
non-perturbative terms share the factorization properties exhibited by
renormalons and (b)
non-perturbative contributions are large in the sense that one can match the
power-like corrections
with the first one or two terms of perturbative expansions.

In the subsequent sections we will try a similar kind of phenomenology on the
$1/Q$ corrections
to the shape variables, in which case there is no OPE but the renormalon
technique is still
available.

\section{Universality of the $1/Q$ corrections to event shapes.}

In this section we summarize and further explore the consequences of the
universality of the $1/Q$
corrections to event shape variables.

In \cite{az} we argued for the universality of the $1/Q$ corrections based
on the following observations:

(a) To all orders in perturbation theory, the $1/Q$ corrections in the
renormalon technique
come from soft gluons.

(b) The soft gluons factorize into a universal factor in the above.
Namely, for an observable
$O$ we can write:
\beq
\langle O \rangle _{1/Q}~=~R_O(\al (Q^2))E_{soft}.
\eeq
Thus, $R_O$ are $O$ dependant and perturbatively calculable whereas $E_{soft}$
are universal and
are given by
\beq
E_{soft}~=~{1\over Q}\int {{d\kp ^2\over \kp^2}\gamma_{eik}(\al (\kp ))\kp }
\eeq
where $\gamma_{eik}$ has been calculated in perturbation theory up to the two
loop
level \cite{kt,rk}. It has also been noted that the same quantity appears in
the
radiative corrections to many hard processes \cite{sterm,trenta,rk}

(c) Combining (a) and (b), together with the discussion of sect. 2, we argue
that since we are
looking at the same $e^+e^-$ annihilation process the above is true
non-peturbatively
as well.

At this point we would like to qualify the above remarks:
The considerations of this paper as well as the earlier one
\cite{az}, apply in the kinematic domain near the semi-inclusive
2 jet limit; i.e. $T \rightarrow 1$. $1/Q$ corrections, in fact,
do not just arise from this domain but also from the multi-jet
kinematic region. Eq (13) is a statement about such corrections
as well and emphasizes that multi-jet configurations are associated
with powers of $\alpha_{s}(Q^2)$. The region $T \rightarrow 1$
itself will be sudakov suppressed at high enough energies and for
such kinematic regimes the multi-jet configurations are important.
In the last section of this paper we return to a discussion of this point.
 For the moment, however, we restrict ourselves only to the region where
the $1/Q$ corrections near the 2-jet limit are important.

We shall estimate the numerical value of $QE_{soft}$, including
non-perturbative
terms, in the next
section and identify the corresponding non-perturbative quantity in terms of
jet masses
in sect. 5.
In the remainder of this section we will discuss further physical quantities ,
observed in
$e^+e^-$ annihilation which also receive $1/Q$ corrections and calculate the
corresponding
$R_O$.

We turn first to a discussion of the jet opening angle. The energy weighted jet
opening angle is an
infrared safe quantity and is defined as \cite{basham}:
\beq
\langle sin^2\eta \rangle ~=~ \langle{\Sigma E_i sin^2\delta_i \over Q}\rangle
\eeq
where $E_i$ and $\delta_i$ are the energy and the angle with respect to the jet
axis of i-th
particle, respectively. Here we would like to demonstrate the existence of
$1/Q$ corrections to this
observable using the renormalon technique.

Consider the contribution to this quantity from the gluon emission process the
crossection for
which is:( $k$ is the gluon momentum and $p_1$ and $p_2$ those of the quark and
antiquark
respectively)
\beq
{1\over{\sigma_{0}}}{d\sigma\over dx_1dx_2}~=~{C_F\over 2\pi} \al
{x_1^2+x_2^2\over (1-x_1)(1-x_2)}
\eeq
where $x_i=2E_i/Q$. Let us change variables from $x_1,x_2$ to $x_2$ and $x=\kp
/Q$
 More explicitly:
\beq
x~=~{\sqrt{(1-x_1)(1-x_2)(x_1+x_2-1)}\over x_2}
\eeq
supposing that the gluon is emitted into the same hemisphere as the
antiquark with energy $E_2$. Then:
\beq
{1\over{\sigma_{0}}}{d\sigma\over dx_2dx}~=~{C_F\over \pi}\al
{(x_1^2+x_2^2)(x_1+x_2-1) \over
(1-x_2)(x_2-2(1-x_1))} {1\over x}
\eeq
with
\beq
x_1~=~1-{x_2\over 2}+{x_2\over 2}\sqrt{1-{4x^2\over {1-x_2}}}
\eeq
and,
\beq
0\le ~x~\le {\sqrt{1-x_2}\over 2}.
\eeq
Anticipating that the $1/Q$ correction comes from the region of soft gluons
$((x_2,x_1\sim1,x_3\rightarrow 0)$
we see that for gluons $E sin^2\delta$ is $O(\omega )$ whereas for the
energetic quantities it is
$O(\omega^2/Q)$. Hence we keep only the contribution of gluons as far as the
$1/Q$  terms are
concerned.

Thus,
\beq
\langle {E sin^2\delta\over Q}\rangle~=~
\int_0^1dx_2\int_0^{1/2\sqrt{1-x_2}}{dx\over x}{2C_F\over \pi}\al
{x^2(x_1^2+x_2^2)(x_1+x_2-1)\over (1-x_2)(x_2-2(1-x_1))(2-x_1-x_2)}.
\eeq
Now to get the renormalon contribution, we led $\al \rightarrow \al (\kp^2)$
\cite{basseto}
and consider the limit $x_1,x_2\sim 1$ and $x$ small. Then:
\beq
\langle {Esin^2\delta\over Q}\rangle ~\approx~{4C_F\over \pi}
\int_0^1dx_2\int_0^{{1\over 2}\sqrt{1-x_2}}dx\cdot x
{\al(\kp ^2)\over (1-x_2)
(1-{x_2\over 2}-{x_2\over 2}\sqrt{1-{4x^2\over 1-x_2}}}.
\eeq
Interchanging the order of integration we get:
\beq
\langle {E sin^2\delta\over Q}\rangle~\approx~
\int_0^1dx\cdot x\int_0^{1-4x^2}dx_2{4C_F\over \pi}{\al(\kp ^2)\over
(1-x_2)^2+x^2}
{}~\approx~{\pi\over 2Q}\int_0^{Q/2}d\kp \cdot {4C_F\over \pi}\al (\kp ).
\eeq
The general features at the origin of these $1/Q$ corrections are
the same as discussed in the previous paper \cite{az}, i.e. they come from a
region of soft gluons
and close to the 2-jet limit.

As final example of the $1/Q$ corrections to jet properties let us consider the
expectation value
of the fractional energy $\eta ~=~2E(\delta)/Q$ emitted inside a double cone of
half opening angle
$\delta$ centered around the thrust axis. This too is an infrared safe quantity
\cite{sw}.
We consider the n-th moment of $\eta$:
\beq
M^n(\delta)~=~\int\eta^n\rho(\eta)d\eta
\eeq
where $\rho(\eta)$ is the probability density describing the emission of the
fractional energy
$\eta$ inside the cone.

We write this moment as \cite{barreiro}:
\beq
M^n(\delta )~=~1-\int(1-\eta^n)\rho(\eta )d\eta ~
\eeq
where the integration region is one with at least one parton outside the cone.
Thus for a
$q\bar{q}g$ final state:
\beq
M^n(\delta)~=~1-{C_F\over 2\pi}\al \int (1-\eta^n)
{x_1^2+x_2^2\over (1-x_1)(1-x_2)}dx_1dx_2.
\eeq
We are interested in determining if there are any power corrections linear in
$Q^{-1}$ to this
quantity. For this, we again concentrate on the region of soft gluons. We note
that in the above
the integration region is subject to the following constraints for the gluon
angle and energy:
\beq
\eta~=~1~if~ sin
\theta<sin\delta;~~~\eta~=~1-x_3~if~sin\theta>sin\delta.\label{constraint}
\eeq
For soft gluons we expand $(1-\eta^n)~\approx~nx_3$ and thus we are let to
consider:
$$
-{nC_F\over 2\pi}\al \int dx_2dx_1{{x_1^2 + x_2^2}\over (1-x_1)(1-x_2)}x_3
 $$
subject to the constraint (\ref{constraint}). In the soft gluon region this
becomes in terms of
the variables $x_2,x$:
\beq
\langle M^n(\delta)\rangle _{1/Q}~=~-n{2C_F\over \pi}
\int dx_2\int {dx\over x}{\al(\kp ^2)\over 1-x_2}\{1-x_2+{x^2\over1-x_2}\}.
\eeq
The integration region is determined from:
\beq
{1\over 2}(1-x_2)sin\delta~\le~x~\le{1\over 2}\sqrt{1-x_2};~~~
0\le~x_2~\le 1.
\eeq
We next interchange the order of integration and concentrate on the small $x$
region to extract
consistently the $1/Q$ term using the renormalon technique. We get:
\begin{eqnarray}
\langle M^n(\delta)\rangle_{1/Q}~\approx~-n{2C_F\over \pi}
\int_0^{{1\over2}sin\delta}{dx\over x}
\int_{1-{2x\over sin\delta }}^{1-4x^2}\al (\kp ^2)
{1\over {1-x_2}}\{1-x_2+{x^2\over 1-x_2}\} \\ \nonumber
=~-n{4C_F\over \pi} {1\over Qsin \delta}
\int_0^{{Q\over 2}sin\delta}d\kp \al (\kp ^2)
-n{2C_F\over \pi}{sin \delta \over Q}\int_0^{{Q\over 2}sin\delta}d\kp \al (\kp
^2).\label{large}
\end{eqnarray}
The conditions for the applicability of this calculation is:
\beq
{n\La \over Q\cdot sin\delta}~\ll ~1.
\eeq
Thus we expect non-perturbative corrections to the quantity $M^n(\delta )$
which for small
$\delta $ goes like $n/Q \delta $. Corrections of the form $1/Q\delta$ for
small
$\delta$ in jet processes was discusssed in \cite{ks1}.

We should again emphasize here that in order to compare these predictions
with the data one has to make sure that the kinematic region under study is
such that the 2-jet configurations are not strongly sudakov suppressed.

\section{Comparison with experimental data.}

In this section we compare the theoretical predictions based on the
universality of
the $1/Q$ terms with experimental data.

At this point, we have four observables $\langle 1-T \rangle , \langle C
\rangle ,
\langle \sigma_L\rangle $ and
$\langle Esin^2\delta /Q\rangle $ on which the idea of the universality of the
$1/Q$ corrections
can be tested experimentally. The results are shown in Table 1 below.
In listing the predictions we have used the following observation consistent
with our approach to renormalon phenomenology discussed in section 2. In
calculating
$<1-T>$, we proceed as in \cite{az} and obtain the corresponding $1/Q$ term
in the lowest non trivial order. To this order this is the same as the $1/Q$
correction to the heavy jet mass, $M_h^2$. Since the effect comes from the
soft region where the coupling is order unit, we must resum all such
contributions.
In higher orders, the light jet mass$M_l^2$ will similarly give a non vanishing
$1/Q$
correction. Thus we must include this in our non perturbative estimate of
$<1-T>$ as well. Consistent with our discussion in section 2 we must therefore
allow for a non perturbative enhancement factor of 2. This is taken into
account
in our predictions for the non perturbative contributions to $<1-T>$ as well as
to $<C>$, and to$<\sigma_{L}/\sigma_{T}>$. We shall come back to a discussion
of this point in sections 4 and 5.

{\centerline \bf Table 1}

\begin{eqnarray*}
Observable & \al~expansion & 1/Q~correction ~(GeV)~~~R_{o}~~~~QE_{soft}~(GeV)
\\
\langle 1-T \rangle & 0.335\alpha +1.02\alpha ^2 &~~~~~~~~~~~~\sim 1
{}~~~~~~~~~~~~~~~~~
2~~~~~~~\sim
0.5 \\  \langle C\rangle & 1.374\al+4\al ^2 &~~~~~~~~~~~~ \sim 5
{}~~~~~~~~~~~~~~~~~3\pi~~ ~~~\sim 0.53
\\ \langle {\sigma_L\over \sigma_T}\rangle &~~~~~~ {\al/ \pi}
&~~~~~~~~~~~~\sim
0.8~~~~~~~~~~~~~~~ {\pi \over 2}~~~~~~\sim 0.51 \\ \langle {Esin^2\delta\over
Q}
\rangle &~~~~~~
{2\al / \pi }& ~~~~~1.20\pm 0.25 ~~~~~~~~~~~~~~\pi~~~~~~0.38\pm 0.08
\end{eqnarray*}
The estimate of the $1/Q$ correction for the first three observables is taken
from \cite{webber}. This paper contains also comparison of theoretical
predictions, in a different form,
with experimental data treating one of variables as an input. In
the next section we shall be able to fix the overall scale as well.
Data on the energy weighted jet angle is from \cite{barreiro}. The
energy range
for the last estimate is also different from the first three: 8-30 GeV for the
latter compared
to upto LEP energies for the former. We should note that using the results of
\cite{barreiro}
for the $1/Q$ correction to thrust, we would find for the quantity
$QE_{soft}=0.30\pm 0.08$.

We did not list prediction (\ref{large}) because it cannot be confronted
directly with experimental fits to $1/Q$ corrections. However, the
data on the moments exist for various values of $\delta, n$
\cite{barreiro}.
An inspection of the corresponding experimental curves reveals that
the data
do not show large non perturbtaive corrections for small $\delta$
as predicted in (\ref{large}). Taken at face value, the data are the most
serious disagreement with theory that we find. Updating of the analysis of the
data appears desirable.

\section{The overall scale of the $1/Q$ corrections.}

In this section we identify the non perturbative scale associated with the
$1/Q$ corrections in $e^{+}e^{-}$ annihilation, i.e we infer the non
perturbative
counterpart of $QE_{soft}$. In order to motivate this identification, we
briefly
recall the discussion of thrust from \cite{az}. There it was shown that to all
orders
the $1/Q$ corrections come from soft gluons and from the neighbourhood of the
two
jet limit. In particular, we identified from simple kinematics and from the
infrared safety of $T$, that to all orders, as far as the $1/Q$ corrections are
concerned:
\beq
\langle 1-T \rangle_{1/Q}= {M_{1}^{2} \over Q^{2}} + {M_{2}^{2} \over Q^{2}}
\eeq
Here we consider two hemispheres divided by a plane perpendicular to the thrust
axis and
$M_{1}^{2}$ and $M_{2}^{2}$ are the invariant masses in these.
In the limit of soft gluons $M_{i}= \sum_{j} 2p_{i}k_{j}$, with $k$ the momenta
of the gluons.Thus in perturbation theory, $\langle 1-T \rangle_{1/Q}$ is
essentially
twice the jet mass squared. By using the factorization of soft gluons, we have
upto
corrections of $O(\alpha_{s}(Q^{2}))$:
\beq
\langle 1-T \rangle_{1/Q}=2 E_{soft}\label{kin}
\eeq
with,
\beq
QE_{soft}|_{renormalon}= \int{dk_{\perp}^{2} \over k_{\perp}^{2}}
\gamma_{eik}(\alpha(k_
{\perp}^{2})) k_{\perp}.
\eeq
where, in the integrand we pick up the contribution of the renormalon pole in
$\alpha_{s}$.

It might worth emphasizing that, once it is established that the $1/Q$
corrections are associated with soft gluons the relation (\ref{kin})
between the jet mass and $<1-T>$ becomes purely kinematical. It is inherent to
hadronization models (for review see \cite{barreiro,webhad}), or to soft
perturbative contributions (see, e.g., \cite{new}).

Now, we use this kinematical nature of relation (\ref{kin}) to fix the
overall
scale of non-perturbative contribution to the $1/Q$ terms.
Namely, we argued in
\cite{az} and in the earlier sections that the
relation which are true to all orders in $\al $ normalized at low mass scale of
$\sim \Lambda_{QCD}$ should be true non perturbatively as well.
Thus, we are naturally led to
the non perturbative identification of $Q^{2} E_{soft}$,
 near the 2-jet limit, as the average non
perturbative correction to the square of the jet masses. Phenomenologically the
hadronization correction to the jet masses is parametrized thus:
\beq
\langle M_{had}^{2} \rangle = \lambda Q
\eeq
which corresponds to jet momenta receiving a negative correction of order
$\lambda$.
For example, in the "tube" model of hadronization one finds ( see \cite{webhad}
for a review),
\beq
\lambda = \int d^{2}p_{\perp} \rho(p_{\perp}) p_{\perp}
\eeq
where $\rho(p_{\perp})$ gives the $p_{\perp}$ distribution of hadrons in a
rapidity-$p_{\perp}$ "tube". In particular, according to the analysis
quoted in \cite{webhad}, from the experimental data on jet masses, one finds:
\beq
\lambda \sim 0.5~ Gev.
\eeq

In fact in simple physical models one can directly see the non perturbative
connection
between the $1/Q$  corrections to the observables and to the quantity
$\lambda$.
Consider for instance the energy weighted jet opening angle
$\langle sin^{2}\eta \rangle$. Let $\tilde{\rho}(z,p_{\perp})$ denote the
appropriately normalized distribution of hadrons in a jet with longitudinal
momentum fraction $z$ and perpendicular component $p_{\perp}$. Thus, if
$p_{3}$ is the component of the hadron momenta along the jet axis, then
$z={2p_{3} / Q}$. Whence,
\beq
\langle sin^{2}\eta \rangle = \int_{0}^{1}dz \int d^{2} p_{\perp}
\tilde{\rho}(z,p_{\perp})
                              {p_{\perp}^{2} \over {1/4 z^{2}Q^{2} +
p_{\perp}^{2}}}
\eeq
The $1/Q$ terms can only come from the neighbourhood of $z=0$, hence:
\beq
\langle sin^{2}\eta \rangle = \int d^{2}p_{\perp}\rho(p_{\perp})\int_{0}^{1}dz
                              {p_{\perp}^{2} \over {1/4 z^{2}Q^{2} +
p_{\perp}^{2}}}
\eeq
where, $\tilde{\rho}(0,p_{\perp})=\rho(p_{\perp})$. Thus,
\begin{eqnarray}
\langle sin^{2}\eta \rangle~ =~ \pi{1 \over Q}\int
d^{2}p_{\perp}\rho(p_{\perp})
p_{\perp} ~                            =~\pi{\lambda \over Q}
\end{eqnarray}
The closeness of the two approaches is hard to miss. The reason for this
similarity
is that both renormalons and the standard hadronization picture identify
bounded intrinsic $p_{\perp}$ as a manifestation of the non perturbative
effects.
The corresponding derivations of the $1/Q$ corrections therefore become
identical
in this approximation, with the replacement
${\gamma_{eik} / k_{\perp}^2} \rightarrow \pi \rho$.( Compare equations
(22) and
(39)).

As a final example consider the moments $M^{n}(\delta)$ discussed in section 3.
The
$( sin\delta)^{-1}$ dependance is easily understood in the hadronization
picture.
Indeed, hadrons with energies upto $p_{\perp} / sin\delta$ leave the cone
with opening angle $\delta$ and reduce the flow of energy inside this cone, in
agreement with the renormalon picture given in section 3. ( See equation (30)).

To summarize, we have proposed to identify the non perturbative scale from
pure kinemtical considerations and the theoretical input on the dominant
contribution of soft gluons to $1/Q$ corrections.So far the absolute scale
of non-perturbative terms was fixed only within the operator product expansion
in the euclidean region \cite{svz}.The generalization of this idea to
Minkowskian space has thus far not produced  a handle on the phenomenology
of $1/Q$ corrections since the operators become nonlocal \cite{ks1,ks2}.
It is amusing that just the same transition from euclidean to minkowskian space
which is responsible for the change of local operators into nonlocal ones
brings new tools based on the minkowskian kinematics. However,
 as emphasized earlier,our identification
can be correct only to order $O(\al(Q^2))$. In the absence of the OPE
the generalization to higher orders in $\al(Q^2)$ is not straightforward at
all,
as we discuss in the next section.

\section{Comparison with other methods. Conclusions.}

In this paper we have attempted to bridge the renormalon
picture
of $1/Q$ corrections with the phenomenological hadronization models. The
considerations
of this and the previous paper \cite{az} may be thought of as providing a field
theoretic argument for the justification, near the 2-jet limit, of these
hadronization
models which were
so far considered at a purely phenomenological level. Conversely, one may turn
the argument around
and use the success of the phenomenological models to justify the picture of
renormalon inspired phenomenology discussed in section 2.

It might be worth mentioning that not all renormalon-based approaches end up
with the same kind of phenomenology as outlined above \footnote{The
following discussion is added following the suggestion of a referee
report.}. Indeed, consider
models with (fictitious) gluon mass $\mu$ \cite{webber} or with freezing of
the coupling constant at a scale $\mu_I$
\cite{dw}. The results based on this model were reviewed
recently in \cite{webrev}. It is no surprise that many
predictions coincide with what we have.
A more detailed consideration unveils however that this coincidence is rather
formal.
The easiest way to appreciate the difference is to consider
predictions for the difference of masses of a heavy and light
jets. According to \cite{webber,dw,webrev}:
\beq
{\langle {M_h^2-M_l^2\over Q^2}\rangle}_{1/Q}~
=~
{\langle {M_h^2 \over Q^2}\rangle}_{1/Q}~=~
{\langle 1-T\rangle}_{1/Q}~=~
{16\al\over 3\pi}{\mu\over Q}~=~{16\over 3\pi}
{\mu_I\over Q}\bar\alpha_0(\mu_I) \label{frozen}
  \eeq
 Note that the prediction for $\langle
1-T \rangle_{1/Q}$ coincides with (\ref{kin}) upon identification
of our $QE_{soft}$ with $(8/3\pi)\mu_I\bar\alpha_0(\mu_I)$. This identification
works in some other cases as well.
However, the central point point about eq (\ref{frozen}) is that adding or
subtracting $M_l^2$ does not change anything.
This($M_{l}^{2}=0$) is
inherent to models with a gluon mass or with
the frozen coupling \cite{webrev}.
The reason is that these models generalize to non-perturbative physics
basing on one-loop calculations.

On the other hand, within the approach developed in this paper
we reproduce in the leading
approximation the standard hadronozation picture as described, say,
by
the tube model. Therefore, the $(M_h^2-M_l^2) $ difference depends
on the distribution of nonperturbative momenta and vanishes if this
distribution is dominated by a characteristic value. It has been known
for some time \cite{barreiro} that this picture is consistent experimentally:
\beq
{\langle{ M_h^2 -M_l^2\over Q^2} \rangle}_{1/Q}~\ll
{\langle {M_h^2\over Q^2} \rangle}_{1/Q}.
\eeq
Similarly, our prediction for the $1/Q$ corrections reads as
\beq
{\langle 1-T \rangle}_{1/Q}~=
{}~{\langle {M_h^2+M_l^2\over Q^2}\rangle}_{1/Q}
{}~\approx~2\cdot {\langle {M_h^2\over Q^2}\rangle}_{1/Q}
\eeq
This difference of a factor of 2 compared to (\ref{frozen}) is not
immediately manifest because of arbitrariness
in normalization of $\mu_I$.

One might wonder how it is possible to have different pictures within the
one and the same renormalon approach. The point is that, as emphasized
in sect. 2, the enhancement factor for nonperturbative effects defies
naive analysis of divergencies of perturbative expansions. In particular,
the model with gluon mass makes non perturbative effects,
as imitated by non analytical terms in mass squared,
 subordinate to the perturbative contributions
in the most transparent way. Indeed, since the $\mu/Q$ correction is associated
with
a small fraction of the phase space this term is always a minor effect
on the background of the ordinary perturbative contribution.
Within this model,
it is inconsistent
to consider both $M_h$ and $M_l$ as nonvanishing unless at least $\al^2$
corrections are included. In fact, more careful analysis would reveal
that even higher orders are needed for consistent treatment of $1/Q$
terms.

Within the aproach which allows for enhancement of non perturbative
effects there is no difficulty to retain $1/Q$ terms even when the
perturbative terms are small numerically. Similarly, in QCD sum rules
the condensate terms dominate over pure perturbative contributions
at moderate $Q^2$ (see sect. 2). It might be worth noting that according to
Ref \cite{webrev} the corrections to the prediction of the model with
the frozen coupling can be as large as $50\%$. In the renormalon language,
our eq (\ref{kin}) corresponds to summation of all these large
corrections so that the remaining uncertainty is of order $\alpha_s (Q^2)$
\cite{az,ks2}. In a sense this corresponds to summing the soft contributions
from the multi-jet configurations as well.
However, this summation cannot be done in the model with
frozen coupling since the freezing has no clear field theoretical realization.

Let us also mention paper \cite{ns} which tests, in particular,the
conclusions of references \cite{az,ks1} on the example of large $n_f$.
Although an agreement is found in lowest orders evaluated
explicitly, the question is raised on possible existence of $\alpha_s(Q^2)
\cdot lnQ^2$ corrections in higher orders, due to multijet events.
It might be worth emphasizing, therefore, that all the statements
on two-jet dominance which are made above are to be qualified for the
energy being not so high as to make the effect of the Sudakov suppression
appreciable. If not,then terms like $lnQ^2\cdot \alpha_s(Q^2)$ do appear.
Moreover, these terms cannot be dealt directly within
the renormalon technique. The way out is well known however.
Namely, one considers the resummed cross section and
renormalon corrections to it \cite{cs,ks1,az,ks2}. For example,
in case of thrust instead of $<1-T>_{1/Q}$ one has to consider \cite{ks2}
\beq
{\langle e^{-\nu(1-T)}\rangle}_{1/Q}
\eeq
where the parameter $\nu$ is large enough to ensure the dominance of
the two-jet configuration despite the Sudakov suppression.
In our discussion of the experimental data we did not introduce $\nu$ since
the fits to $1/Q$ corrections \cite{webber,dw} do not indicate any
energy dependence.

Another issue raised in Ref \cite{ns} is the search for a proper choice
of a
shape variable which would be free of $1/Q$ corrections and hence
could be used for reliable extractions of the value of $\alpha_s$.
We would like to notice that consideration of spherocity could be
promising from
this point of view. The spherocity   $S$ is defined as \cite{gm}
\beq
S~=~\left({4\over \pi}\right)^2~min
\left( {\Sigma_i|p_T^i|\over \Sigma_i|P_i|}\right)^2.
\eeq
Perturbatively, it is given as \cite{derujula}
\beq
\langle S\rangle~=~{64\over \pi^2}{\alpha_s\over 3\pi}\left( -{229\over
9}+64ln{3\over 2}\right).
\eeq
Explicit calculation shows that the spherocity is free from
$1/Q$ correction in the leading order in $\alpha_s$. Thus to this accuracy:
\beq
\langle S\rangle ~=~ {\alpha_s\over 3\pi}(3.28)~+~ O(1/Q^2)
\eeq
Instread of giving details of derivation let us remind the reader of the simple
estimate of the non perturbative contribution to the spherocity
\cite{derujula}:
\beq
\Delta S_{non-pert}~\approx~
\left({4\over \pi}\right)^2{<p_T>^2\over Q^2} <n(Q^2)>^2~\sim
{}~{lnQ^2\over Q^2}
\eeq
where $n(Q)$ is the multiplicity and $<p_T>$ is the characteristic non
perturbative transverse momentum. Since we know that renormalons
do reproduce the basic features of the phenomenological hadronization models,
it is no surprise that that there is no $1/Q$ renormalon here. Moreover,
this kind of argument appears to work for multijet events. Thus, there are
good chances that the spherocity is indeed free from $1/Q$ corrections
to higher orders in $\alpha_s(Q)$ as well.

As we emphasized above,
all of our considerations until now have only dealt with the leading $1/Q$
corrections to the quark jets. For gluon jets, the corresponding anomalous
dimension $\gamma_{eik}$ is different in perturbation theory. Thus it appears
likely that a different non perturbative parameter will be needed to describe
situations dealing with gluon jets. Continuing in a similar spirit,we recall
that
our estimates of the $1/Q$ corrections have been carried out in the kinematic
region
where the dominant contribution comes from the neighbourhood of the 2-jet
limit.
 However it is clear that
though suppressed, three jet processes will have to be taken into account in
the next
order of approximation,i.e order $\alpha_{s}(Q^{2})$. New non perturbative
parameters
could arise here. It is a challenge to implement such corrections into the
developed
framework.

\section{Acknowledgements}

We are grateful to F. Barreiro and R. Peschanski for discussions. RA would like
to thank the
Service Physique Theorique, CE Saclay, for hospitality and  support and VIZ
acknowledges the
hospitality of Saclay and IPN, Orsay. This work was supported in part by the US
Department
of Energy.


\begin{thebibliography}{99}
\bibitem{troyan}Yu.L. Dokshitzer, V.A. Khoze, A.H. Mueller and S.I. Troyan,
"Basic of Perturbative QCD", Editions Frontieres, Paris, (1991);
G. Sterman, et. al. Rev. Mod. Phys., {\bf 67} (1995) 157.
\bibitem{collins}J.C. Collins, D. Soper
and G. Sterman, {\it in "Perturbative QCD"}, A.H. Mueller Editor, World
Scientific, (1989).
\bibitem{thrust}Z. Kunst, P. Nason, G. Marchesini and
B.R. Webber, QCD: {\it in } Proc. of the Workshop on Z physics  at LEP, eds. G.
Altarelli, R. Kleiss
and C. Verzegnassi, CERN Report 83-08 (1989).
\bibitem{barreiro}F. Barreiro, Fortshr. Phys., {\bf 34} (1986) 503.
\bibitem{webhad}B.R. Webber, Proc. Summer School on Hadronic Aspects of
Collider Physics,
Zuoz, Switzerland, August 1994, ed. M.P. Locher (PSI, Villigen (1994)).
\bibitem{svz}M.A.Shifman, A.I. Vainshtein, and V.I. Zakharov, Nucl. Phys.,
{\bf B147} (1979)  385, 447.
\bibitem{cs}H. Contopanagos and G. Sterman, Nucl.Phys., {\bf B419} (1994) 77.
\bibitem{webber}B.R. Webber, Phys. Lett., {\bf B339} (1994) 148;
\bibitem{ks1}G. Korchemsky and G. Sterman, Nucl.Phys., {\bf B437} (1995) 415.
\bibitem{dw}Yu.L. Dokshitzer and B.R. Webber, Phys.Lett. {\bf B352} (1995) 451,
hep-ph/9504219.
\bibitem{az}R. Akhoury and V.I. Zakharov,  Phys.Lett.,{\bf B357} (1995) 646.
\bibitem{ks2}G. Korchemsky and G. Sterman, hep-ph/9505391
\bibitem{bigi}I.I. Bigi, M.A. Shifman, N.G. Uraltsev, and A.I. Vainshtein,
Phys. Rev. {\bf D50}
(1994) 2234; M. Beneke and V.M. Braun, Nucl. Phys. {\bf B426} (1994) 301.
\bibitem{mw}A.V. Manohar and M.B. Wise, Phys. Lett. {\bf B344} (1995) 407
\bibitem{thooft}G. 't Hooft, {\it in} Whys of Subnuclear Physics, Erice 1977,
Ed. A. Zichichi, Plenum NY, (1977) p. 943;
B. Lautrup, Phys. Lett. {\bf B69} (1977) 109; G. Parisi, Phys.Lett. {\bf B76}
(1978) 65.
\bibitem{mueller1}A.H.Mueller, Phys. Lett., {\bf B308} (1993) 355.
\bibitem{bbz}M. Beneke, V.M. Braun, and V.I. Zakharov, Phys. Rev. Lett. {\bf
74} (1994) 3057.
\bibitem{david}F. David, Nucl.Phys., {\bf B234} (1984) 237; Nucl.Phys., {\bf
B263} (1986) 637.
\bibitem{mueller2}A.H. Mueller, Nucl.Phys., {\bf B250} (1985) 327.
\bibitem{novikov}V.A. Novikov, M.A. Shifman, A.I. Vainshtein and V.I. Zakharov,
Nucl.Phys., {\bf B237} (1984) 525.
\bibitem{vz}V.I. Zakharov, Nucl.Phys., {\bf B385} (1992) 452.
\bibitem{basham}C.L. Basham, L.S. Brown, S.D. Ellis and S. T. Love, Phys. Rev.
{\bf D19} (1979) 2018
\bibitem{basseto}D. Amati, A. Bassetto, M. Ciafaloni, G. Marchesini and G.
Veneziano,
Nucl. Phys. {\bf B173} (1980) 429
\bibitem{sw}G. Sterman, S.Weinberg, Phys. Rev. Lett., {\bf 39} (1977) 1436
\bibitem{kt}J. Kodaira, L.Trentadue, Phys. Lett {\bf B112} (1982) 66
\bibitem{rk}G. P. Korchemsky, A. V. Radyushkin, Phys. Lett. {\bf B171} (1986)
459;
Nucl. Phys. {\bf B283} (1987) 342
\bibitem{sterm}G. Sterman,Nucl.Phys. {\bf B281} (1987) 310
\bibitem{trenta}S. Catani and L. Trentadue, Phys. Lett. {\bf B217} (1989) 539;
 Nucl.Phys. {\bf B327} (1989)  323;
 Nucl. Phys. {\bf B353} (1991) 183.
\bibitem{new}S. Catani, L. Trentadue, G. Turnock and B. Webber, Nucl.Phys.
{\bf B407} (1993) 7.
\bibitem{webrev}B.R. Webber, hep-ph/9510283.
\bibitem{ns}P. Nason and M.H. Seymour, Nucl. Phys.,{\bf B454} (1995) 291.
\bibitem{gm}H. Georgi and M. Machacek, Phys. Rev. Lett.,{\bf 39} (1977) 1237.
\bibitem{derujula}A. De Rujula, J. Ellis, E.G. Floratos and M.K. Gaillard,
Nucl. Phys. {\bf B138} (1978) 387.

\end{thebibliography}
\end{document}